\documentclass[12pt]{article}
\usepackage[utf8]{inputenc}
\usepackage{comment,times}
\usepackage{xurl}
\usepackage{hyperref}
\usepackage{graphicx}
\usepackage[table,xcdraw]{xcolor}
\usepackage[caption=false]{subfig}
\usepackage{caption}
\usepackage{enumitem}
\usepackage{float}
\usepackage{longtable}
\usepackage{setspace}
\usepackage[skip=20pt plus1pt, indent=20pt]{parskip}
\usepackage{atbegshi}
\usepackage{booktabs}

\usepackage{algorithm}
\usepackage[noend]{algpseudocode}
\let\oldReturn\Return
\renewcommand{\Return}{\State\oldReturn}
\algrenewcommand\algorithmicrequire{\textbf{Precondition:}}
\algrenewcommand\algorithmicensure{\textbf{Postcondition:}}

\title{
  Enhancing Efficiency in Parallel Louvain Algorithm for Community Detection
  \vspace{-2ex}
}
\author{
  Subhajit Sahu \\
  Center for Security, Theory, and Algorithmic Research (CSTAR) \\
  IIIT Hyderabad, India - 500 032 \\
  subhajit.sahu@research.iiit.ac.in
}
\date{}

\begin{document}
\maketitle

\begin{abstract}
Community detection is a key aspect of network analysis, as it allows for the identification of groups and patterns within a network. With the ever-increasing size of networks, it is crucial to have fast algorithms to analyze them efficiently. It is a modularity-based greedy algorithm that divides a network into disconnected communities better over several iterations. Even in big, dense networks, it is renowned for establishing high-quality communities. However it can be at least a factor of ten slower than community discovery techniques that rely on label-propagation, which are generally extremely fast but obtain communities of lower quality. The researchers have suggested a number of methods for parallelizing and improving the Louvain algorithm. To decide which strategy is generally the best fit and which parameter values produce the highest performance without compromising community quality, it is critical to assess the performance and accuracy of these existing approaches. As we implement the single-threaded and multi-threaded versions of the static Louvain algorithm in this report, we carefully examine the method's specifics, make the required tweaks and optimizations, and determine the right parameter values. The tolerance between each pass can be changed to adjust the method's performance. With an initial tolerance of 0.01 and a tolerance decline factor of 10, an asynchronous version of the algorithm produced the best results. Generally speaking, according to our findings, the approach is not well suited for shared-memory parallelism; however, one potential workaround is to break the graph into manageable chunks that can be independently executed and then merged back together.
\end{abstract}

\section{Introduction}
The proliferation of interconnected data from real-world sources, such as social and biological networks, has led to an increase in the use of graphs as a means of representation \cite{graph-sakr21}. These graphs, however, are often massive in size, necessitating the use of parallelism to handle the scale of the data. Additionally, many real-world graphs are dynamic, with edges being constantly added and removed \cite{com-zarayeneh21}. As a result, research into parallel algorithms for analyzing and updating graph analytics on dynamic graphs has gained significant attention in recent years. Examples of this research include the dynamic calculation of centrality scores \cite{cent-lerman10, cent-li10, cent-bergamini16, cent-nathan17, cent-regunta21}, maintenance of biconnected components \cite{conn-galil91, conn-liang01, conn-haryan22}, and computation of shortest paths \cite{path-narvaez00, path-chan08, path-khanda21}.

Community detection is a widely studied problem in graph analysis, with practical applications in fields such as e-commerce, communication networks, and healthcare. It involves identifying groups of vertices in a graph, known as communities, that are densely connected within the group but sparsely connected to the rest of the network. When these structures can be identified based solely on the topology of the network, they are referred to as intrinsic communities. On the other hand, extrinsic communities are defined based on external information or attributes of the nodes, such as membership in a particular organization or geographic location. There are various types of communities that can be identified in a graph. Disjoint communities are those in which each vertex belongs to exactly one community (studied here). Alternatively, overlapping communities \cite{com-gregory10} allow each vertex to belong to more than one community, while hierarchical communities have a multi-level membership structure.

Community detection is a challenging problem as it is NP-hard and the number of communities and their size distribution is not known in advance. There are various techniques for addressing this problem, such as label propagation \cite{com-raghavan07, com-gregory10, com-xie11}, random walk \cite{com-rosvall08}, diffusion \cite{com-kloster14}, spin dynamics \cite{com-reichardt06}, fitness metric optimization \cite{com-newman06, com-fortunato10}, statistical inference \cite{com-come15, com-newman16}, simulated annealing \cite{com-guimera05, com-reichardt06}, clique percolation \cite{com-derenyi05, com-maity14, com-gupta22}, and more. These methods can be grouped into two main categories: divisive and agglomerative. Divisive methods, also called top-down methods, start by assuming all vertices in a graph belong to one community and iteratively identify and remove bridges to split into more well-connected communities \cite{com-girvan02, com-souravlas21}. Agglomerative methods, or bottom-up methods, merge two or more communities together such that a certain score is maximized \cite{com-zakrzewska15, com-zarayeneh21}. Another approach is seed set expansion, which begins with a set of relevant seed vertices of interest, and expands to form communities surrounding them \cite{com-whang13}.

When using different methods for community detection, the communities that are returned can be evaluated using a quality function. One popular measure is modularity, introduced by Newman and Girvan, which compares the number of edges within a community to the expected number in a random-null model. It ranges from -0.5 (non-modular clustering) to 1.0 (fully modular clustering) and optimizes this function theoretically results in the best possible grouping. Another fitness score is conductance, which measures the community cut or inter-community edges. Another popular quality function is the Constant Potts Model (CPM), which aims to overcome some limitations of modularity \cite{com-traag11}.

\section{Literature survey}
There are several techniques that have been developed for detecting communities in networks. A number of them are based on modularity-optimization, hierarchical clustering, label propagation, region density, core clustering \cite{com-ruan15}, game theory, information theory (infomap) \cite{com-zeng19, com-zeng18, com-faysal19, infomap-rosvall09, com-rita20}, and biological evolution (genetics) \cite{com-taufan20, com-ghoshal19, com-lu20}. Metrics such as the modularity score \cite{com-newman06, com-blondel08, com-ghoshal19}, Normalized Mutual Information index (NMI) \cite{com-jain17, com-chopade17}, and Jaccard Index \cite{com-jain17} are used to compare the quality of communities obtained with different approaches.

The \verb|Louvain| method is a greedy modularity-based optimization algorithm that hierarchically agglomerates vertices in a graph to obtain communities. It was created by Blondel et al. \cite{com-blondel08} from the University of Louvain, and is one of the most popular heuristics in this field. It has an average time complexity of $\Theta (n \log n)$, with $n$ being the total number of nodes in the network \cite{com-lancichinetti09}. Approaches to perform the Louvain algorithm can be either \textbf{coarse-grained}, where a set of vertices are processed in parallel; or fine-grained, where all the vertices are processed in parallel. Several parallelization heuristics for the Louvain algorithm have been implemented in Grappolo software library \cite{com-halappanavar17}. It should be noted though that community detection methods such as the Louvain that rely on modularity maximization are known to suffer from \textbf{resolution limit problem}. This prevents identification of communities of certain sizes \cite{com-ghosh19}. Some \textbf{improvements on the Louvain algorithm} include using a suitable heuristic based partitioning \cite{com-zeng15}, dealing with ghost vertices between graph partitions \cite{com-zeng15},  restricting the internal search rules \cite{com-ryu16}, and early pruning of the non-promising candidates \cite{com-ryu16}. Other interesting approaches include the use of MapReduce in a BigData batch processing framework \cite{com-zeitz17}.

One of the main advantages of the Louvain algorithm is its ability to find communities with high modularity, which is a measure of the density of connections within communities and the sparsity of connections between communities. However, the Louvain algorithm does have some limitations. It can be sensitive to the initial order of the nodes, which can lead to non-reproducible results. It can also have difficulty detecting communities that are very dense or tightly interconnected, as it tends to break up dense communities in favor of finding larger, more cohesive communities.

\section{Evaluation}
In this section, we first describe our experimental setup, such as the system we use and our dataset. We then investigate the static Louvain algorithm, taking note of the subtle details and explore various optimizations, while implementing their single-threaded and multi-threaded OpenMP-based versions.

\subsection{Experimental setup}

\subsubsection{System used}

In our experiments, we employed a system comprised of two Intel Xeon Silver 4116 64-bit processors running at 2.10 GHz, and 128GB of DDR4 Synchronous Registered DRAM operating at 2666 MHz. Each processor featured 12 x86 cores, each with 2 hyper-threads per core, and 16.5M L3 cache. Our server was running the CentOS version 7.9, with GCC version 9.3 and OpenMP version 5.0 used for compilation with optimization level 3 (-O3) and simultaneous multi-threading (SMT) was enabled for all experiments.

\subsubsection{Dataset}

The graphs used in our experiments are detailed in Table \ref{tab:dataset}. These graphs were obtained from the SuiteSparse Matrix Collection \cite{suite19}. The total number of vertices in the graphs varies from $74.9$ thousand to 12 million, and the total number of edges varies from $811$ thousand to $304$ million. All edges are considered to be undirected and weighted with a default weight of one, and self-loops were added to each vertex in all the graphs.

\begin{table}[!ht]
\centering
\caption{In our experiments, we use a list of 17 graphs. Each graph has its edges duplicated in the reverse direction to make them undirected, and a weight of 1 is assigned to each edge. The table lists the total number of vertices ($|V|$), total number of edges ($|E|$) after making the graph undirected, and the average degree of vertices ($D_{avg}$) for each graph. The number of vertices and edges are rounded to the nearest thousand or million, as appropriate.}
\label{tab:dataset}
\begin{tabular}{||c||c|c|c|c||}
  \toprule
  \textbf{Graph} &
  \textbf{$|V|$} &
  \textbf{$|E|$} &
  \textbf{$D_{avg}$} \\
  \midrule
  \multicolumn{4}{|c|}{Web Graphs} \\ \hline
    web-Stanford & 282K & 3.99M & 14.1 \\ \hline
    web-BerkStan & 685K & 13.3M & 19.4 \\ \hline
    web-Google & 916K & 8.64M & 9.43 \\ \hline
    web-NotreDame & 326K & 2.21M & 6.78 \\ \hline
    indochina-2004 & 7.41M & 304M & 41.0 \\ \hline
    \multicolumn{4}{|c|}{Social Networks} \\ \hline
    soc-Slashdot0811 & 77.4K & 1.02M & 13.2 \\ \hline
    soc-Slashdot0902 & 82.2K & 1.09M & 13.3 \\ \hline
    soc-Epinions1 & 75.9K & 811K & 10.7 \\ \hline
    soc-LiveJournal1 & 4.85M & 86.2M & 17.8 \\ \hline
    \multicolumn{4}{|c|}{Collaboration Networks} \\ \hline
    coAuthorsDBLP & 299K & 1.96M & 6.56 \\ \hline
    coAuthorsCiteseer & 227K & 1.63M & 7.18 \\ \hline
    coPapersCiteseer & 434K & 32.1M & 74.0 \\ \hline
    coPapersDBLP & 540K & 30.5M & 56.5 \\ \hline
    \multicolumn{4}{|c|}{Road Networks} \\ \hline
    italy\_osm & 6.69M & 14.0M & 2.09 \\ \hline
    great-britain\_osm & 7.73M & 16.3M & 2.11 \\ \hline
    germany\_osm & 11.5M & 24.7M & 2.15 \\ \hline
    asia\_osm & 12.0M & 25.4M & 2.12 \\ \hline
  \bottomrule
\end{tabular}
\end{table}

\subsection{Louvain algorithm}

\textbf{Louvain algorithm}, as mentioned before, is an agglomerative-hierarchical community detection method that greedily optimizes for modularity (iteratively). Given an undirected weighted graph, all vertices are first considered to be their own communities. In the first phase, also known as the \textbf{local-moving phase}, each vertex greedily decides to move to the community of one of its neighbors which gives the greatest increase in modularity. If moving to no neighbor's community leads to an increase in modularity, the vertex chooses to stay with its own community. This is done sequentially for all the vertices. If the total change in modularity is more than a certain threshold (\verb|tolerance| parameter), this phase is repeated. Once this phase is complete, all vertices have formed their first hierarchy of communities. The next phase is called the \textbf{aggregation phase}, where all the vertices belonging to a community are collapsed into a single super-vertex, such that edges between communities are represented as edges between respective super-vertices (edge weights are combined), and edges within each community are represented as self-loops in respective super-vertices (again, edge weights are combined). Together, the \textit{local-moving} and the \textit{aggregation phases} constitute a \textbf{pass}. This \textit{super-vertex graph} is then used as input for the next pass. This process continues until the increase in modularity is below a certain threshold (\verb|pass_tolerance| parameter, which is generally $0$ as we want to maximize our modularity gain). As a result from each pass, we have a hierarchy of community memberships for each vertex as a \textit{dendrogram} \cite{com-leskovec21}. We generally consider the \textit{top-level hierarchy} as the final result of the community detection process.

Adjusting \verb|tolerance| between each pass (known as \textit{threshold scaling}) has been observed to impact runtime of the algorithm, without significantly affecting the modularity of obtained communities. We conduct experiments to obtain a suitable rate at which \textit{tolerance} can be decreased between each pass (\verb|tolerance_decline_factor| parameter), in addition to the initial \verb|tolerance| parameter value that would be suitable on average for most graphs. In our first experiment, we implement a \textbf{single-threaded CPU-based version} of the Louvain algorithm. We adjust the initial value of \verb|tolerance| from $1$ to $10^{-12}$ in steps of $10$, and adjust the tolerance\_decline\_factor from $10$ to $10^4$. From the results, we observe that an initial \verb|tolerance| of $0.01$ yields communities with the best possible modularity while requiring the least computation time. In addition, increasing the \verb|tolerance_decline_factor| increases the computation time (as one might expect), but does not seem to impact resulting modularity. Thus, a \verb|tolerance_decline_factor| of $10$ would be good.

Authors of the original paper use an \textbf{asynchronous} version of the algorithm, where the community membership of each vertex can be dependent upon the community membership of its neighbors in the current iteration (similar to \textit{Gauss-Seidel} method) \cite{com-blondel08}. Anyhow, it suffers from reads and writes to the same memory area, which can be \textit{detrimental} to performance in a \textit{multi-threaded implementation} due to cache coherence overhead. We therefore consider experimenting with a \textbf{synchronous} version of the algorithm, where the community membership of each vertex can only be dependent upon the community membership of its neighbors in the previous iteration (similar to \textit{Jacobi} method). We perform this comparison on the \textit{single-threaded CPU-based implementation} of the algorithm. From the results, we observe that both the synchronous and asynchronous version of the algorithm are able to provide communities of \textit{equivalent quality} in terms of \textit{modularity}, with the asynchronous version providing slightly higher quality communities for certain graphs. However, the \textit{synchronous version} is quite a bit slower than the asynchronous one in terms of the total \textit{time} taken, as well as the total number of \textit{iterations} of the local-moving phase (which is the most expensive part of the algorithm). We therefore conclude that \textit{asynchronous / partially asynchronous approaches for vertex processing} are likely to provide \textit{good performance} over fully synchronous versions for parallel implementations of the Louvain algorithm. Partially asynchronous vertex ordering via graph coloring has been explored by Halappanavar et al \cite{com-halappanavar17}.

For the second experiment, we implement a \textbf{multi-threaded OpenMP-based version} of the Louvain algorithm. Similar to earlier algorithms, each thread is given a \textit{separate hashtable}, which it can use for choosing to move to a community with the highest delta-modularity. If multiple communities yield the highest delta-modularity, we pick only the first one in the hashtable. As before, hashtables are \textit{allocated separately} for better performance (instead of storing them contiguously on a vector). We use an OpenMP schedule of ``\verb|auto|'' and a total of $12$ threads for the time being (not an optimal choice). We adjust the number of threads from $2$ to $48$ threads on a \textit{dual-socket system} with each CPU having $12$ \textit{cores each} and $2$ \textit{hyperthreads} per core.

From the results, we observe that increasing the number of threads only decreases the runtime of the Louvain algorithm by a small amount. This indicates that when multiple reader threads and a writer thread access common memory locations (here it is the community membership of each vertex) performance is degraded (likely due to higher pressure on cache coherence), and would tend to approach the performance of a sequential algorithm if there is just too much overlap. Utilizing \textit{all} $48$ threads for community detection significantly increases the time required for obtaining the results, and is likely due to thread switching with the operating system. The number of iterations required to converge also increases with the number of threads, indicating that the behavior of the asynchronous multi-threaded implementation starts to approach the behavior of a synchronous version of the algorithm, which converges much more slowly than the asynchronous version. One approach to resolve these issues could be to \textit{partition} the graph in such a way that each partition can be run \textit{independently}, and then combined back together.

\section{Conclusion}
The Louvain algorithm is well-liked among researchers because of its ability to locate high-quality communities in networks. Our research focuses on the static Louvain algorithm. We pay close attention to the algorithm's minute details when implementing its single-threaded and multi-threaded OpenMP-based variants, making any necessary adjustments or optimizations, and obtaining appropriate parameter values. By altering the \verb|tolerance| between each pass, also known as \textit{threshold scaling}, the method's performance can be modified. According to our findings, communities with high modularity are produced while using the least amount of computing time when an \textit{asynchronous} version of the algorithm is used, with an initial \verb|tolerance| of $0.01$ and a \verb|tolerance_decline_factor| of $10$. A parallel OpenMP-based implementation of the algorithm suggests that the algorithm is \textit{not} generally well-suited for shared-memory \textit{parallelism} (unless the input graph has a large number of vertices). A possible solution can be to divide the graph into manageable \textit{partitions} that can each be independently run and then reassembled.

\appendix
\small
\bibliographystyle{IEEEtran}
\bibliography{main}
\end{document}